\documentclass{PoS}
\usepackage{epsfig}
\usepackage{bm}
\usepackage{amssymb}
\usepackage{fontenc}
\usepackage{times}
\usepackage{mathptmx}
\usepackage{graphicx}

\title{Applying Complex Langevin to Lattice QCD at finite $\mu$.}

\ShortTitle{Applying Complex Langevin to Lattice QCD at finite $\mu$.}

\author{\speaker{D.~K.~Sinclair}
        \thanks{This research was supported in part by US Department of Energy
         contract DE-AC02-06CH11357}\\
HEP Division, Argonne National Laboratory, 9700 South Cass Avenue, Argonne, 
Illinois 60439, USA\\
E-mail: \email{dks@anl.gov}}

\author{J.~B.~Kogut\\
Department of Energy, Division of High Energy Physics, Washington, DC 20585,
USA\\
and\\
Department of Physics -- TQHN, University of Maryland, 82 Regents Drive, 
College Park, MD 20742, USA\\
        E-mail: \email{jbkogut@umd.edu}}

\abstract{We continue our simulations of lattice QCD at finite quark-number 
chemical potential, $\mu$,  using the complex-Langevin equation (CLE) with
gauge-cooling and adaptive updating. The CLE is used because QCD at finite
finite $\mu$ has a complex fermion determinant, which prevents use of standard
simulation methods. Simulations using the standard lattice action show a
transition from hadronic to nuclear matter for $\mu < m_\pi/2$ rather than the
expected $\mu \approx m_N/3$. This suggests that the CLE is being influenced
by the phase-quenched theory, which has a transition at $\mu = m_\pi/2$. We
are therefore performing CLE simulations with a new action which includes an
irrelevant chiral 4-fermion interaction. This separates the physics at
energies of order of the pion mass and smaller from that at energies of the
other hadrons. In doing this, it breaks the extended symmetry of the
phase-quenched theory over that of the full theory, raising the masses of the
extra pion-like excitations consisting of a quark and a conjugate quark, which
could otherwise produce such an anomalous transition. Our preliminary CLE
simulations using massless quarks, so that $m_\pi=0$, show no transition at
$\mu=m_\pi/2=0$, but do show a transition at an appreciably higher value of
$\mu$. It remains to be seen if this transition is near to $m_N/3$.}

\FullConference{37th International Symposium on Lattice Field Theory - Lattice2019\\
		16-22 June 2019\\
		Wuhan, China}

\begin{document}

\section{Introduction}

QCD at finite quark-/baryon-number describes nuclear matter. Because QCD at
finite quark-number density has a sign problem, standard methods of simulating
it on the lattice, which are based on importance sampling, fail. When one
implements finite density by using a quark-number chemical potential, $\mu$,
the sign problem manifests itself as a complex fermion determinant. One
promising simulation method which can cope with complex `probabilities' is
the complex-Langevin equation (CLE) 
\cite{Parisi:1984cs,Klauder:1983nn,Klauder:1983zm,Klauder:1983sp}
However, the validity of this method can only be proved if the drift term in
the CLE is holomorphic in the fields, and the domain over which the fields
vary is exponentially bounded. For (lattice) QCD at finite $\mu$, zeros of the
fermion determinant give poles in the drift term, so the drift term is
meromorphic, not holomorphic
\cite{Aarts:2009uq,Aarts:2011ax,Nagata:2015uga,Nishimura:2015pba,Nagata:2016vkn,
Aarts:2017vrv,Seiler:2017wvd,Aarts:2017hqp,Nagata:2018net}.
Hence careful testing of the CLE is needed to determine over what if any range
of parameters, the CLE produces correct results. The CLE has been used to
study lattice QCD at finite $\mu$ in the heavy-dense limit
\cite{Aarts:2008rr,Aarts:2013uxa,Aarts:2014bwa,Aarts:2016qrv,Langelage:2014vpa,
Rindlisbacher:2015pea}. 
Less extensive studies have been made of full lattice QCD at finite $\mu$
\cite{Sexty:2013ica,Aarts:2014bwa,Fodor:2015doa,Nagata:2016mmh,Tsutsui:2018jva,
Scherzer:2018udt,Ito:2018jpo,Tsutsui,Sexty:2019vqx}.

We have performed CLE simulations of 2-flavour lattice QCD at $\beta=6/g^2=5.6$,
$5.7$ \cite{Kogut:2019qmi}
and some preliminary simulations at $\beta=5.8$. Although these indicate
that, for sufficiently weak coupling, correct results are obtained for 
$\mu << m_\pi/2$ and for $\mu$ large enough to produce saturation, these
simulations produce incorrect results for intermediate $\mu$ values. In 
particular, they show a transition from hadronic to nuclear matter at
$\mu < m_\pi/2$ instead of the predicted $\mu \approx m_N/3$. While it is 
possible that the correct physics might be obtained for sufficiently weak
coupling, we are testing modifying the lattice action as a way of producing more
physical results. 

The action we choose is the standard unimproved staggered-quark action with
an additional chiral 4-fermion interaction which preserves the $U(1)$ chiral
symmetry of the original action. Since this additional term is an irrelevant
operator it should not change continuum physics. This term separates the low
energy physics associated with the pion mass from the high energy physics
associated with the other hadrons. QCD can then be described in terms of pions,
heavy `constituent' quarks and gluons rather than in terms of light `current'
quarks and gluons. With this action, the additional chiral symmetry breaking
of the phase-quenched theory is broken, driving the extra pion-like excitations
of a quark and a conjugate quark to higher mass. Our preliminary CLE simulations
with this action and massless quarks at finite $\mu$ show no transition at
$\mu=m_\pi/2=0$, but do show a transition at a considerably higher $\mu$. It 
will require simulations of the phase-quenched theory based on this action to
decide whether this improved behaviour indicates that this is the expected
transition from hadronic to nuclear matter, or whether it is a transition
associated with the new phase-quenched theory at half the mass of its heavy
pion-like excitations with quark-number $2$.

\section{Lattice QCD with a chiral 4-fermion interaction at finite $\mu$}

After introducing auxiliary fields $\sigma$ and $\pi$, our modified Euclidean
action for staggered fermions is:
\begin{equation}
{\cal L}=\frac{1}{4}F_{\mu\nu}F_{\mu\nu}
 +\bar{\psi}(D\!\!\!\!\!/+\gamma_4\mu+\sigma+i\gamma_5\tau_3\pi+m)\psi
 +\frac{\gamma N_f}{8}(\sigma^2+\pi^2).
\end{equation}
Such theories have been studied at $\mu=0$ \cite{Kogut:1998rg}, and for
$\mu \neq 0$ \cite{Barbour:1996pn,Chavel:1997ec}.

After integrating out the fermion fields, the lattice action is
\begin{equation}
S=\beta\sum_\Box \left\{1-\frac{1}{6}Tr[UUUU+(UUUU)^{-1}]\right\}
 -\frac{N_f}{4}{\rm Tr}\{ln[M(U,\mu,\sigma,\pi)]\}
 +\frac{N_f\gamma}{8}\sum_{\tilde{s}}(\sigma^2+\pi^2)
\end{equation}
where $(\tilde{s})$ refers to sites on the dual lattice and the Dirac operator
$M$ is
\begin{equation}
M = D(U,\mu) + \frac{1}{16} \sum_i [\sigma_i+i\epsilon(n)\pi_i] + m
\end{equation}
where $D(U,\mu)$ is the staggered $D\!\!\!\!\!\!/$ in the presence of
quark-number chemical potential $\mu$ and $i$ runs over the 16 sites on the
dual lattice adjacent site $n$ and backward links are represented by $U^{-1}$.
We shall henceforth refer to this theory as chiral QCD, $\chi$QCD.

We note that this lattice action has a $U(1)$ chiral symmetry at $m=0$. The
action is invariant under the global chiral transformation
\begin{equation}
\sigma_i+i\pi_i \rightarrow e^{i\phi}[\sigma_i+i\pi_i]
\end{equation}
whence
\begin{equation}
\sigma_i+i\epsilon(n)\pi_i \rightarrow e^{i\phi\epsilon(n)}
[\sigma_i+i\epsilon(n)\pi_i].
\end{equation}
Since the Jacobian of this transformation is $1$, the theory is invariant
under this transformation. Hence, when $\langle\sigma\rangle$ is non-zero,
$\pi$ is a massless Goldstone boson.

The complex Langevin equations are:
\begin{equation}
-i \left(\frac{d}{dt}U_l\right)U_l^{-1} = -i \frac{\delta}{\delta U_l}
S(U,\sigma,\pi)+\eta_l
\end{equation}
where $\eta_l=\eta_l^\alpha\lambda^\alpha$, and
\begin{eqnarray}
\frac{d\sigma_i}{dt} &=&
-\frac{\delta}{\delta\sigma_i}S(U,\sigma,\pi)+\eta^\sigma_i \\ \nonumber
\frac{d\pi_i}{dt} &=& -\frac{\delta}{\delta\pi_i}S(U,\sigma,\pi)+\eta^\pi_i
\end{eqnarray}
$\eta_l^\alpha$, $\eta^\sigma_i$, $\eta^\pi_i$ are independent gaussian random
numbers appropriately normalized. We discretize this CLE as we did for the
standard action, applying gauge-cooling \cite{Seiler:2012wz} and adaptive 
updating.

\section{Simulations of lattice $\chi$QCD at zero temperature}

We simulate 2-flavour ($N_f=2$) $\chi$QCD with $\beta=5.6$, $\gamma=5$, $m=0$
on a $16^4$ lattice at finite $\mu$. At $\mu=0$ and for $\mu$ large enough to
produce saturation, this lattice is large enough that the theory is in the
confined phase, and at a reasonable approximation to zero temperature. We are
currently performing simulations at a selection of intermediate $\mu$ values
on a $16^3 \times 36$ lattice to check that $16^4$ is adequate to approximate
zero temperature. Note, the addition of the 4-fermion interaction does allow
us to simulate at $m=0$ where we know that $m_\pi=0$. $\beta=6/g^2=5.6$
represents a moderate gauge coupling. While $\gamma=5$ represents a relatively
large 4-fermion coupling ($\gamma$ is inversely proportional to the 4-fermion
coupling), it is not small enough to produce chiral-symmetry breaking
($\chi$SB) without the gluon interactions. $\gamma=5$ was chosen since it is
small enough to allow us to distinguish $\chi$SB at $m=0$ on a $16^4$ lattice.

To date, we have performed CLE simulations of $\chi$QCD with these parameters
for $0 \le \mu \le 0.6$, performing runs of $2 \times 10^6$ updates per $\mu$,
except close to the transition where we performed $3 \times 10^6$ updates per
$\mu$. We see evidence for a phase(?) transition for $\mu \approx 0.35$. We
also performed a short simulation at $\mu=1.5$ where we observe saturation.

First we examine the 2 order parameters which measure the chiral condensate.
These are $\langle\bar{\psi}\psi\rangle$ and $\langle\sigma\rangle$. These
are not independent, but are related by
\begin{equation}
\langle\bar{\psi}\psi\rangle = \gamma\langle\sigma\rangle.
\label{eqn:chiral}
\end{equation}
Because we are simulating at zero quark mass, the direction of the symmetry
breaking in the $(\sigma,\pi)$ or $(\bar{\psi}\psi,\bar{\psi}\gamma_5\xi_5\psi)$
plane is arbitrary. In fact, because we use a finite lattice, this direction
rotates during the simulation, which is how the fact that there is no 
spontaneous symmetry breaking on a finite lattice is enforced. We therefore
use the replacements
\begin{equation}
\sigma = \sqrt{ [{\rm real}(\sigma)]^2+[{\rm real}(\pi)]^2 }
\end{equation}
and 
\begin{equation}
\bar{\psi}\psi = \sqrt{ [{\rm real}(\bar{\psi}\psi)]^2
        +  [{\rm real}(i\bar{\psi}\gamma_5\xi_5\psi)]^2 },
\end{equation}
which approach the actual chiral condensates in the large lattice limit. Note
$\sigma$, $\pi$, $\bar{\psi}\psi$ and $i\bar{\psi}\gamma_5\xi_5\psi$ are lattice
(but not ensemble) averages. We note that now, because of fluctuations, 
equation~\ref{eqn:chiral} is only true in the infinite volume limit. Note that
the imaginary parts of $\sigma$, $\pi$, $\bar{\psi}\psi$ and
$i\bar{\psi}\gamma_5\xi_5\psi$ are very small and have been neglected.

\begin{figure}[htb]  
\parbox{2.9in}{      
\epsfxsize=2.9in     
\epsffile{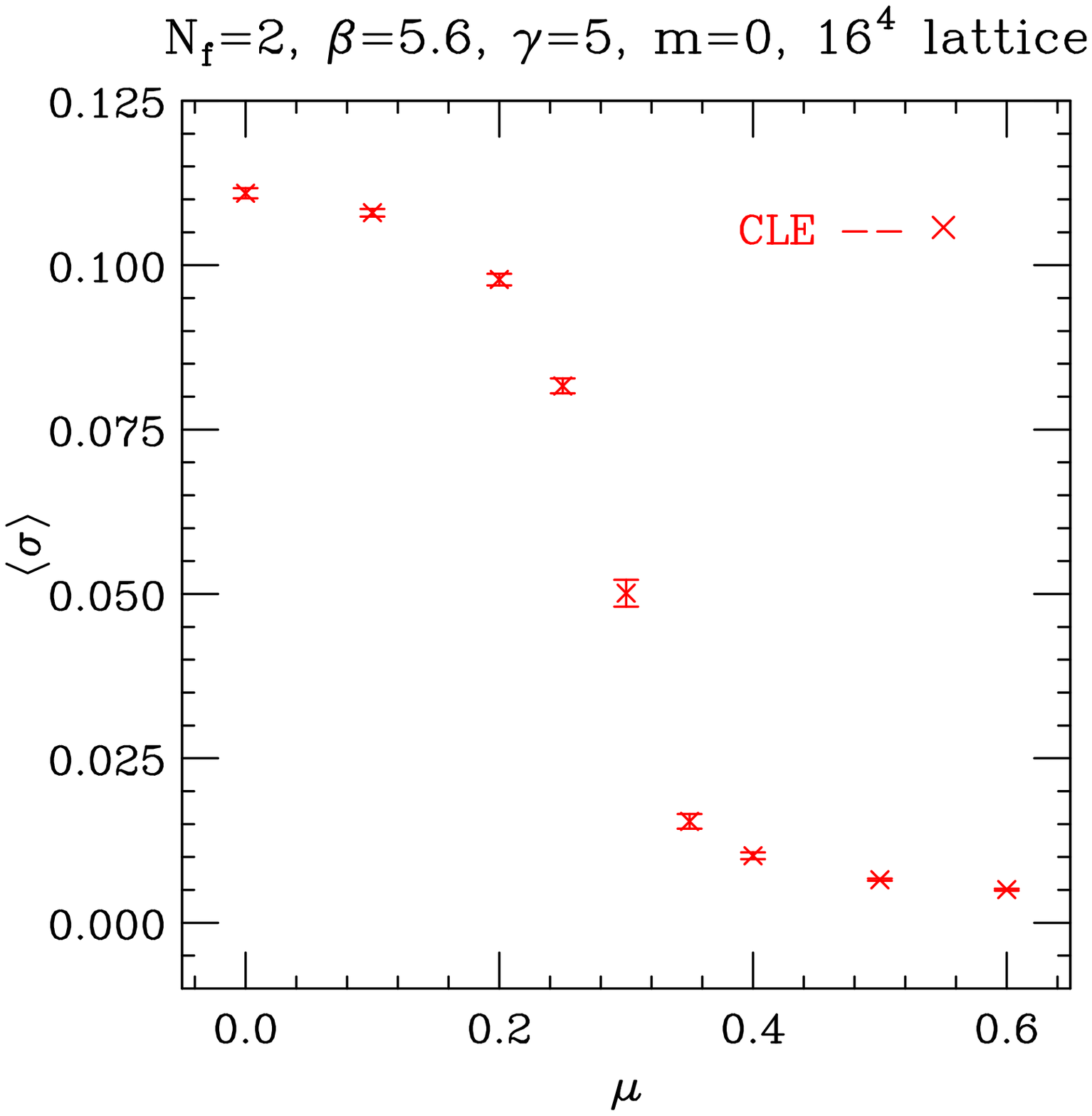}
}                  
\parbox{0.2in}{}               
\parbox{2.9in}{                   
\epsfxsize=2.9in               
\epsffile{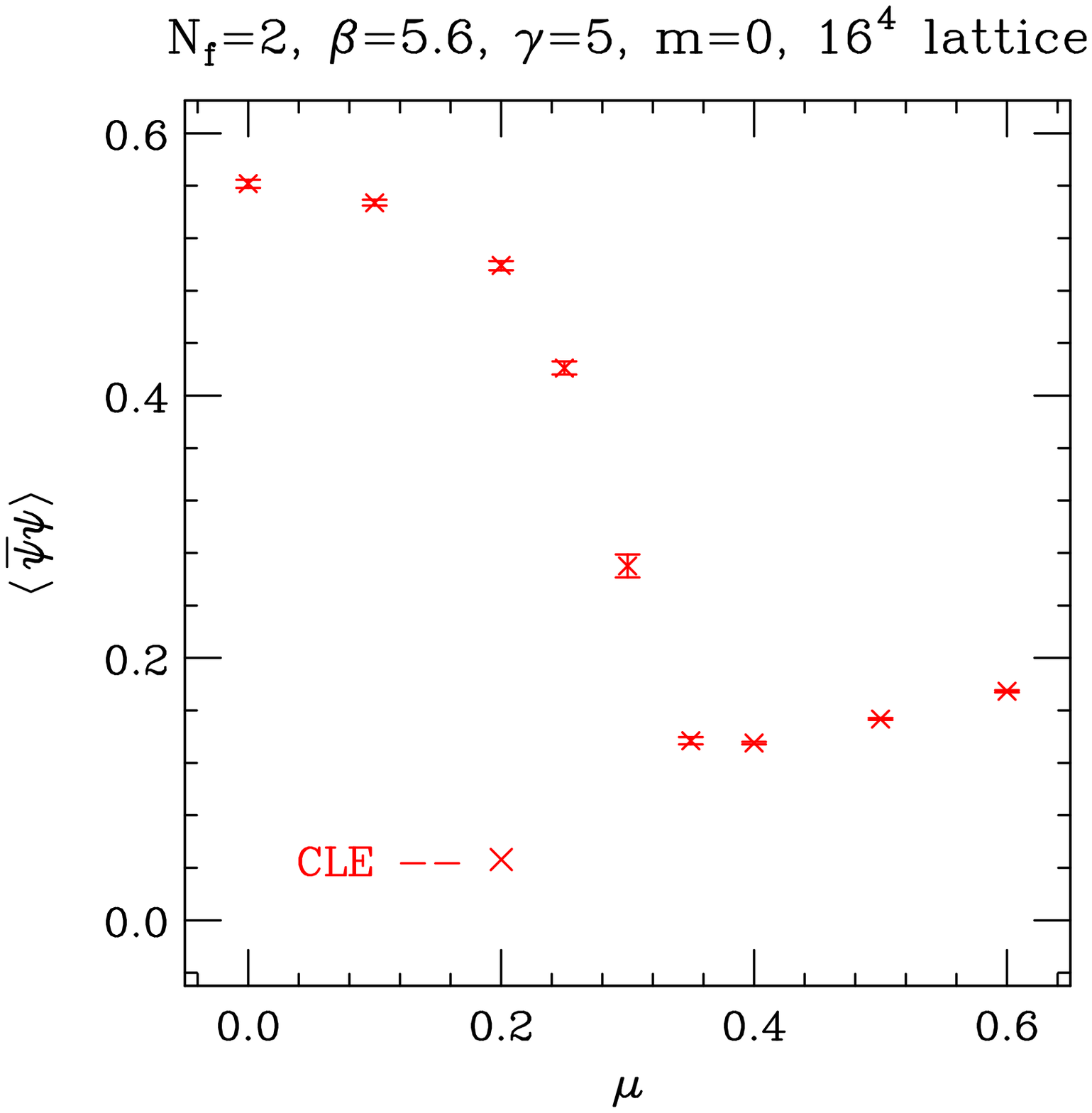}
}                             
\caption{Chiral condensates $\langle\sigma\rangle$ and 
$\langle\bar{\psi}\psi\rangle$ as functions of $\mu$.}
\label{fig:condensates}
\end{figure}

Figure~\ref{fig:condensates} shows the chiral condensates $\langle\sigma\rangle$
and $\langle\bar{\psi}\psi\rangle$ as functions of $\mu$ obtained from our CLE
simulations. We observe that there is no sign of a transition at 
$\mu=m_\pi/2=0$. There is, however, a clear sign of a phase transition at
$\mu \approx 0.35$. Therefore the addition of the 4-fermion term has removed
or moved the transition at an anomalously small $\mu$. This suggests that
the transition at $\mu \le m_\pi/2$, seen in CLE simulations of the original
action does indicate that the CLE is being influenced by the phase-quenched
theory with its superfluid transition at $\mu=m_\pi/2$. It remains to be seen
whether the transition with this new action is the expected hadronic- to
nuclear-matter transition driven by nucleons, or the transition of the new
phase-quenched theory driven by the condensation of the now-heavy
quark-conjugate quark pion-like states of that theory. To test this will
require simulating the new phase-quenched theory, and performing spectroscopy
with the new action and its phase-quenched counterpart. We note that the
chiral condensates do not remain constant up to the transition but rather fall
smoothly once $\mu > 0$. It remains to be seen if this falloff slows at weaker
couplings as is the case with the standard action.

\begin{figure}[htb]
\parbox{2.9in}{
\epsfxsize=2.9in
\epsffile{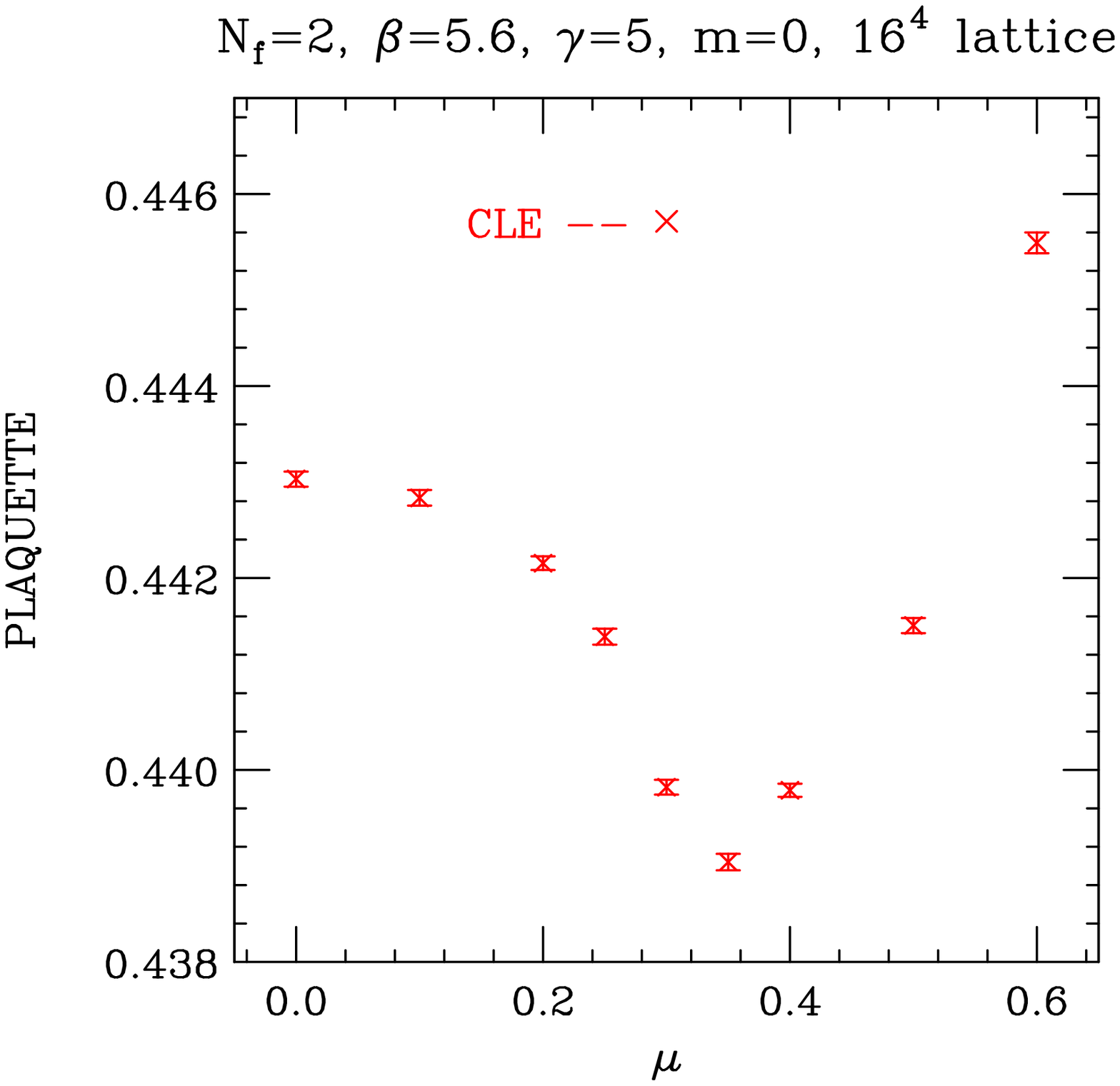}
\caption{Average plaquette as a function of $\mu$.}
\label{fig:apq}
}
\parbox{0.2in}{}
\parbox{2.9in}{
\epsfxsize=2.9in
\epsffile{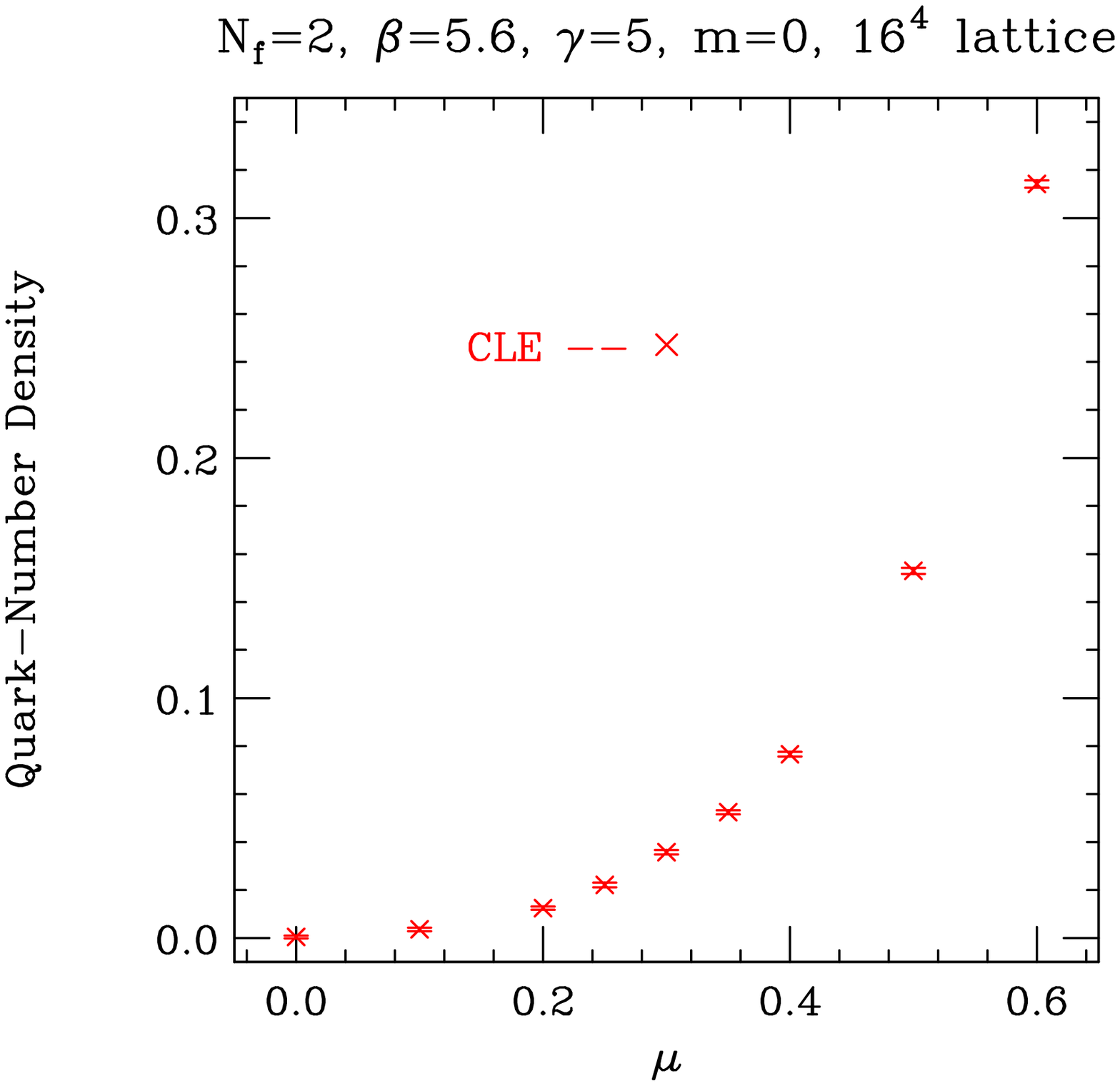}
\caption{Quark-number density as a function of $\mu$.}
\label{fig:qnd}
}
\end{figure}

Figure~\ref{fig:apq} shows the average plaquette 
(Plaquette~$=1-\frac{1}{6}Tr[UUUU+(UUUU)^{-1}]$) as a function of $\mu$. The
most striking feature is the sharp minimum at $\mu \approx 0.35$, consistent
with the position of the transition in the chiral condensates. 
Figure~\ref{fig:qnd} shows the quark-number density as a function of $\mu$.
It appears to be a smoothly increasing function of $\mu$; the rate of increase
increases with $\mu$. For large enough $\mu$ we know that it approaches 
saturation where the quark-number density is $3$, indicating that all fermion 
states are filled and the quarks decouple from the gauge fields.

\section{Summary, discussion and conclusions}

Application of the CLE to simulating lattice QCD at finite $\mu$ using the
standard staggered-quark lattice QCD action predicts a transition from
hadronic to nuclear matter at a $\mu < m_\pi/2$ rather than at the expected
$\mu \approx m_N/3$. This suggests that the CLE is influenced by the
phase-quenched theory with its superfluid transition at $\mu=m_\pi/2$, as is
observed in random matrix models \cite{Mollgaard:2013qra,Bloch:2017sex}.
(Note: there are indications that gauge-cooling might help solve this problem
for random matrix models \cite{Nagata:2016alq}.) We are therefore performing
CLE simulations using an action incorporating a chiral 4-fermion interaction
($\chi$QCD), which explicitly breaks the additional chiral symmetry of the
phase-quenched theory, forcing the masses $m_\Pi$ of the pion-like
excitations, which give rise to the superfluid transition, to larger values.

Our CLE simulations of $\chi$QCD with $m=0$ so that $m_\pi=0$ show no transition
at $\mu=m_\pi/2=0$, but show strong evidence for a phase transition at a higher
$\mu$ value. It remains to be seen if this transition is at $\mu \approx m_N/3$,
or at the new superfluid transition of the new phase-quenched theory at
$\mu = m_\Pi/2$. This will require simulations of phase-quenched $\chi$QCD,
and spectrum calculations for $\chi$QCD and phase-quenched $\chi$QCD at $\mu=0$.

We will need to simulate at weaker gauge and 4-fermion coupling to see if this
improvement will survive to the continuum limit. In addition we will need to
look for evidence that the $\mu$ dependence of observables below the
transition weakens with decreasing coupling. Eventually we will need to provide
4-fermion couplings with the full $SU(2) \times SU(2)$ chiral symmetry.

Other attempts to remedy the problems of applying the CLE to simulations of
lattice QCD at finite $\mu$ involve either adding additional relevant
operators \cite{Nagata:2018mkb} to the QCD action which improve the behaviour
of the CLE and taking the limit as these extra operators vanish, or modifying
the dynamics by adding irrelevant terms to the drift term
\cite{Attanasio:2018rtq}. Since these irrelevant terms have no domain of
holomorphicity, one must take the limit as these extra terms vanish.

\end{document}